\documentclass[showpacs,a4paper,12pt,pra,onecolumn]{revtex4}%
\usepackage{amsfonts}
\usepackage{amsmath}
\usepackage{amssymb}
\usepackage{graphicx}%
\usepackage{epstopdf}
\providecommand{\U}[1]{\protect\rule{.1in}{.1in}}
\ifx\pdfoutput\relax\let\pdfoutput=\undefined\fi
\newcount\msipdfoutput
\ifx\pdfoutput\undefined\else
\ifcase\pdfoutput\else
\ifx\paperwidth\undefined\else
\ifdim\paperheight=0pt\relax\else\pdfpageheight\paperheight\fi
\ifdim\paperwidth=0pt\relax\else\pdfpagewidth\paperwidth\fi
\fi\fi\fi
\begin{document}
\title{An exact master equation for the system-reservoir dynamics under the strong
coupling regime and non-Markovian dynamics}
\author{T. B. Batalh\~{a}o$^{1,2}$, G. D. de Moraes Neto$^{2}$, M. A. de Ponte$^{3}$,
and M. H. Y. Moussa$^{2}$}
\affiliation{$^{1}$Centro de Ci\^{e}ncias Naturais e Humanas, Universidade Federal do ABC,
09210-170, Santo Andr\'{e}, S\~{a}o Paulo, Brazil.}
\affiliation{$^{2}$Instituto de F\'{\i}sica de S\~{a}o Carlos, Universidade de S\~{a}o
Paulo, 13560-970, S\~{a}o Carlos, S\~{a}o Paulo, Brazil.}
\affiliation{$^{3}$Universidade Regional do Cariri, Departamento de F\'{\i}sica,
BR-63010970, Juazeiro Do Norte, CE, Brazil.}

\begin{abstract}
In this paper we present a method to derive an exact master equation for a
bosonic system coupled to a set of other bosonic systems, which plays the role
of the reservoir, under the strong coupling regime, i.e., without resorting
to{ either} the rotating-wave or secular approximations.{ Working with
phase-space distribution functions, }we verify that the dynamics {are
separated in the evolution of its center, which follows classical mechanics,
and its shape, which becomes distorted. This is the generalization of a result
by Glauber, who stated that coherent states remain coherent under certain
circumstances, specifically when the rotating-wave approximation and a
zero-temperature reservoir are used. We show that the counter-rotating terms
generate fluctuations that distort the vacuum state, much the same as thermal
fluctuations. Finally, we discuss conditions for non-Markovian dynamics. }

\end{abstract}

\maketitle

\section{Introduction}

The subject of open quantum systems has undergone substantial growth in the
last three decades, starting with contributions to the field of fundamental
quantum physics with the aim of understanding the process of decoherence.
Based on the von Neumann approach to the reduction of the state vector
\cite{Neumann}, these contributions were mainly driven by the pioneering work
of Zurek \cite{Zurek}, Caldeira and Leggett \cite{CL}, and Joos and Zeh
\cite{JZ}. The repercussions of their work, together with the advent of the
field of quantum information theory, led to renewed interest in open quantum
systems, the focus now shifting from fundamental issues to practical
applications in circuits to implement quantum logic operations.

The master equation approach has long been used to derive system-reservoir
dynamics, to account for energy loss under a weak coupling regime
\cite{Walls}. Its effectiveness comes from the fact that the energy loss of
most quantum mechanical systems, especially within quantum and atomic optics,
can be handled by the single-pole Wigner-Weisskopf approximation \cite{WW},
where a perturbative expansion is performed in the system-reservoir coupling.
Following developments by Caldeira and Leggett \cite{CL}, more sophisticated
methods to deal with the system-reservoir strong coupling regime have been
advanced, such as the Hu-Paz-Zhang \cite{HPZ} master equation, with
time-dependent coefficients, which allows for non-Markovian dynamics.
Halliwell and Yu \cite{HY} have published an alternative derivation of the
Hu-Paz-Zhang equation, in which the dynamics is represented by the Wigner
function, and an exact solution of this equation was given by Ford and
O'Connell \cite{FO}.

Recently, the non-Markovian dynamics of open quantum systems has been studied
with renewed interest, especially in connection with quantum information
theory, as in Refs. \cite{Nori,Wu}. However, in these studies, as well as in
most of the derivations of master equations with time-dependent coefficients,
the authors assume either the rotating-wave approximation (RWA) or the secular
approximation (SA) for the system-reservoir coupling \cite{Makela}. Since
non-Markovian behavior is sensitive to the counter-rotating terms in the
interaction Hamiltonian, important features of the dynamics are missing under
the RWA in the strong-coupling regime. It is worth mentioning that a study of
the effect of the RWA and the SA on the non-Markovian behavior in the
spin-boson model at zero temperature has already been advanced \cite{Makela},
without, however, deriving a master equation.

Our goal in this work is to derive {and investigate the consequences of} a
master equation within the strong-coupling regime, which prevents us resorting
to either the RWA or the SA in the system-reservoir coupling. Moreover,
instead of the path integrals approach \cite{FH}, we use the formalism of
quasi-probability distributions, thus enabling us to cast the problem as the
solution of a linear system of equations. Our results follow from the general
treatment of a bosonic dissipative network we have previously presented in
Ref. \cite{MickelGeral}, where the network dynamics were investigated, and
further used for quantum information purposes \cite{MickelBunch}. However,
differently from our previous developments, we first consider the general
model for a network of bosonic non-dissipative oscillators and, subsequently,
we focus on some of these oscillators (or in just one of them) as our system
of interest, and treat all the others as a (structured) reservoir. The exact
dynamics of the network allows us to obtain an exact dynamics of the
system-reservoir interaction. Moreover, we present a simple inequality to
distinguish between Markovian and non-Markovian dynamics.

Finally, this development enables us to generalize an earlier result by
Glauber \cite{GlauberBook}.{ When using the RWA and a zero-temperature
reservoir, it was shown that the quasi-probability functions maintain their
shape while they are displaced in phase space; in particular, coherent states
remain coherent states}. We find that, for a general Gaussian state, the
center of its phase space distribution follows classical dynamics (as in Ref.
\cite{GlauberBook}), but its shape is changed. Furthermore, this change can be
derived from the evolution of the vacuum state, which is no longer stationary,
because of the counter-rotating terms. The change in shape is affected by both
quantum and thermal fluctuations, and these contributions can be
distinguished, at least in theory. Our developments can be straightforwardly
translated to the derivation of an exact master equation for fermionic
systems, using the reasoning in Ref. \cite{Glauber}.

\section{Unitary dynamics of the universe}

\label{sec:model}

\begin{figure}[ptb]
\begin{center}
\includegraphics[
height=8.0cm,
width=10.0cm
]%
{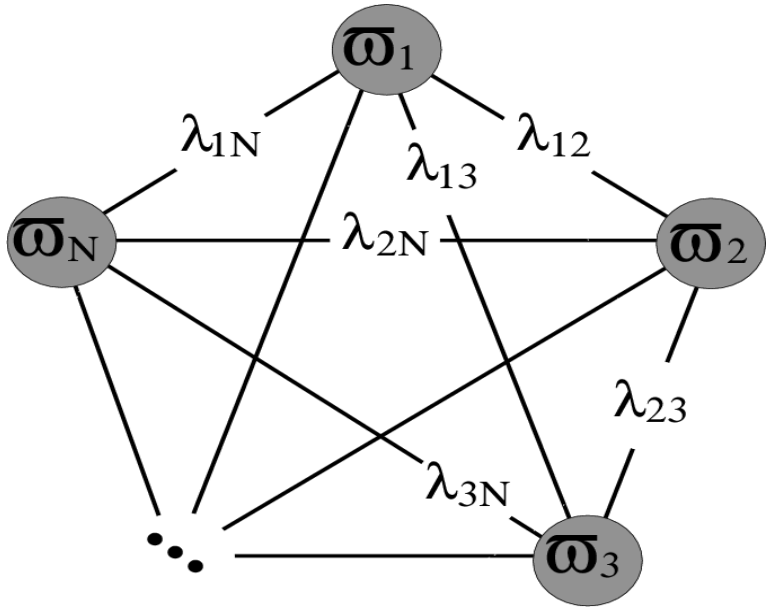}%
\caption{Network of
coupled quantum harmonic oscillators in a general topology.}%
\end{center}
\label{fig:fig1}%
\end{figure}

The universe considered here consists of a set of $M+N$ harmonic oscillators,
which are linearly coupled to each other in an arbitrary network. We consider
$M$ of them to be part of our system of interest, and the remaining $N$ to be
part of a reservoir. However, at this stage, we are concerned with the full
dynamics of the universe, and there is actually no difference between system
and reservoir modes. The oscillators are described by mass $m_{k}$ and
natural, isolated frequencies $\varpi_{k}$; the coupling between modes $k$ and
$j$, which occurs via their position coordinates, has strength $\lambda_{kj}$
(which, without loss of generality, is symmetric in its indices). Before we
write the Hamiltonian that describes such a universe, we note that it must be
positive-definite, in order to be bounded from below and have a well-defined
ground state. Then, the Hamiltonian which is compatible with this model is
\begin{equation}
H=\frac{1}{2}\sum_{k=1}^{M+N}\left(  \frac{1}{m_{k}}\hat{p}_{k}^{2}%
+m_{k}\varpi_{k}^{2}\hat{q}_{k}^{2}\right)  +\frac{1}{4}\sum_{kj=1}%
^{M+N}\lambda_{kj}\left(  \hat{q}_{k}-\hat{q}_{j}\right)  ^{2},
\label{eq:hamiltonqp}%
\end{equation}
where t{he coefficients $\lambda_{kj}$ form a real, symmetric matrix. We do
not assume any particular form for them, so as to generate an arbitrary
network, as depicted in Fig. \ref{fig:fig1} }The coupling term induces a
change in the natural frequency of each mode, that is now represented by
\begin{equation}
\omega_{k}=\sqrt{\varpi_{k}^{2}+\frac{1}{m_{k}}\sum_{j=1}^{N}\lambda_{kj}}.
\end{equation}

Using this renormalized frequency, we can define annihilation operators
$a_{k}$ and rewrite the Hamiltonian as
\begin{equation}
H=\sum_{k=1}^{M+N}\omega_{k}a_{k}^{\dagger}a_{k}+\frac{1}{2}\sum_{kj=1}%
^{M+N}g_{kj}\left(  a_{k}+a_{k}^{\dagger}\right)  \left(  a_{j}+a_{j}%
^{\dagger}\right)  , \label{eq:hamiltona}%
\end{equation}
the coupling in this picture being given by
\begin{equation}
g_{kj}=\frac{\lambda_{kj}}{2\sqrt{m_{k}m_{j}\omega_{k}\omega_{j}}}.
\label{eq:grenorm}%
\end{equation}

From here on, we will focus on $\omega_{k}$ and $g_{kj}$, the latter forming a
real, symmetric matrix.

\subsection{Characteristic function}

The dynamics given by the Hamiltonian of Eq. (\ref{eq:hamiltona}) is best
understood in terms of the characteristic function of a state, which is just
the expected value of the multimode displacement operator in the symmetric
ordering,{
\begin{equation}
\chi\left(  \left\{  \beta_{k}\right\}  \right)  =\left\langle \prod
_{k=1}^{M+N}\exp\left(  \beta_{k}a_{k}^{\dagger}-\beta_{k}^{\ast}a_{k}\right)
\right\rangle \;,
\end{equation}
where $\left\{  \beta_{k}\right\}  $ represents all coordinates $\beta_{k}$
with $k=1,\dots,N$, as well as their complex conjugates. }

The characteristic function carries the complete information about the state,
and in particular information about moments of all orders; this is one of the
reasons it is a better approach than using the Heisenberg equations of motion
directly. The von Neumann equation in Hilbert space is mapped to a
differential equation in dual phase space (where the characteristic function
is defined):%
\begin{equation}
\frac{\partial\chi}{\partial t}=i\sum_{k=1}^{M+N}\left(  \omega_{k}\beta
_{k}-\sum_{j=1}^{N}g_{kj}\left(  \beta_{j}+\beta_{j}^{\ast}\right)  \right)
\frac{\partial\chi}{\partial\beta_{k}}+\text{ H.c.}.
\end{equation}

Being linear and of first order, this equation admits a simple ansatz,
\begin{equation}
\chi\left(  \left\{  \beta_{k}\right\}  ,t\right)  =\chi\left(  \left\{
\beta_{k}\left(  t\right)  \right\}  ,0\right)  , \label{eq:ansatz}%
\end{equation}
which implies that the characteristic function maintains its shape, but the
underlying (dual) phase space undergoes a linear transformation, given by
\begin{equation}
\beta_{k}\left(  t\right)  =\sum_{j=1}^{M+N}\left(  U_{j,k}\left(  t\right)
\beta_{j}-V_{j,k}\left(  t\right)  \beta_{j}^{\ast}\right)  .
\label{eq:linear}%
\end{equation}
This transformation is defined by the solution to a system of differential
equations,
\begin{subequations}
\begin{align}
\frac{dU_{kj}}{dt}  &  =i\omega_{j}U_{kj}-i\sum_{n=1}^{M+N}\left(
U_{k,n}-V_{k,n}\right)  g_{n,j},\label{s1}\\
\frac{dV_{kj}}{dt}  &  =-i\omega_{j}V_{kj}-i\sum_{n=1}^{M+N}\left(
U_{k,n}-V_{k,n}\right)  g_{n,j}. \label{s2}%
\end{align}
The Heisenberg equations of motion for the first moments have a similar
structure. However, since they refer only to first moments, they do not
represent a complete solution of the problem, which can be obtained from the
characteristic function with the same computational effort.

\section{Reduced dynamics of the system}

From this point on, we shall be interested only in the behavior of a subset of
$M$ oscillators (the ones labeled $1$ to $M$), which form our system of
interest, while the oscillators labeled $M+1$ to $M+N$ play the role of a
(structured) reservoir. The complete solution to the dynamics is given by
Eq.(\ref{eq:ansatz}); in order to eliminate the reservoir degrees of freedom,
all we need to do is set $\beta_{k}=0$ if $k>M$ (i.e., evaluate the
characteristic function at the origin of the phase space of the modes we want
to eliminate from the description). Before continuing, we observe that
although not strictly necessary in our method, for the sake of simplicity we
assume the usual sudden-coupling hypothesis, i.e., that the states of system
and reservoir are initially uncorrelated:
\end{subequations}
\begin{equation}
\chi_{SR}\left(  \left\{  \beta_{k}\right\}  ,0\right)  =\chi_{S}\left(
\left\{  \beta_{k}\right\}  _{k\leq M},0\right)  \chi_{R}\left(  \left\{
\beta_{m}\right\}  _{m>M}\right)  . \label{eq:initial}%
\end{equation}
Tracing out the reservoir degrees of freedom, following the procedure above,
leads to
\begin{equation}
\chi_{S}\left(  \left\{  \beta_{k}\right\}  ,t\right)  =\chi_{S}\left(
\left\{  \beta_{k}\left(  t\right)  \right\}  ,0\right)  \chi_{\text{in}%
}\left(  \left\{  \beta_{k}\right\}  ,t\right)  \;, \label{eq:reducedsolution}%
\end{equation}
where the indices run only through the degrees of freedom of the system (i.e.,
$k$ runs from $1$ to $M$). Therefore, we must use Eq.(\ref{eq:linear}) with
$\beta_{k}=0$ for $k>M$, and it follows that we only need $U_{kj}$ and
$V_{kj}$ for $k\leq M$. Eqs. (\ref{s1},\ref{s2}), although written as a matrix
equation, are actually a set of $N$ independent vector equations and we
conclude that only a few of these need to be solved. In fact, if our system of
interest were a single oscillator, we would reduce the problem of finding its
exact dynamics to a single vector equation of dimension $2N$.

The two terms of Eq. (\ref{eq:reducedsolution}) are called the homogeneous
(because it depends on the initial state of the system) and inhomogeneous
terms (because it is independent of it, depending only on the initial state of
the reservoir). The homogeneous part of the solution is just the linear
transformation of phase space induced only by the elements $U_{kj}$ and
$V_{kj}$ for which both $k,j\leq M$. These elements can be arranged in two
general complex $M\times M$ matrices, resulting in $4M^{2}$ real parameters.

At this point, we make an additional assumption that the initial state of the
reservoir is Gaussian \cite{Gaussian}, i.e., its characteristic function has
the Gaussian form. Moreover, the reservoir is unbiased (i.e., $\left\langle
a_{m}\right\rangle =0$ for $m>M$). These are reasonable hypotheses, since the
Gaussian states include the thermal states of quadratic Hamiltonians. The
inhomogeneous characteristic function is then also a Gaussian function:
\begin{align}
\chi_{in}\left(  \left\{  \beta_{k}\right\}  ,t\right)   &  =\exp\left(
-\frac{1}{2}\sum_{kj=1}^{M}A_{kj}\left(  t\right)  \beta_{k}\beta_{j}^{\ast
}\right) \nonumber\\
&  \times\exp\left(  \sum_{kj=1}^{M}B_{kj}\left(  t\right)  \beta_{k}\beta
_{j}+\text{c.c}\right)  \text{.}%
\end{align}
The time-dependent functions $A_{kj}$ and $B_{kj}$ may be divided into two
terms, in the form $A_{kj}=A_{kj}^{\left(  0\right)  }+A_{kj}^{\left(
th\right)  }$ (and similarly for $B$), the first of which is the solution for
a zero-temperature reservoir,
\begin{subequations}
\label{eq:pqzero}%
\begin{align}
A_{kj}^{\left(  0\right)  }  &  =\frac{1}{2}\sum_{m=M+1}^{M+N}\left(
U_{km}U_{jm}^{\ast}+V_{km}V_{jm}^{\ast}\right) \\
B_{kj}^{\left(  0\right)  }  &  =\frac{1}{2}\sum_{m=M+1}^{M+N}\left(
U_{km}V_{jm}+V_{km}U_{jm}\right)  \;,
\end{align}
while the second incorporates the effects of the reservoir initial state,
which is completely characterized by the second-order moments $\left\langle
a_{m}^{\dagger}a_{n}\right\rangle _{0}$ and $\left\langle a_{m}a_{n}%
\right\rangle _{0}$,
\end{subequations}
\begin{subequations}
\label{eq:pqtemp}%
\begin{align}
A_{kj}^{\left(  th\right)  }=  &  \sum_{m=M+1}^{M+N}\left\langle
a_{m}^{\dagger}a_{n}\right\rangle _{0}\left(  U_{km}U_{jn}^{\ast}+V_{kn}%
V_{jm}^{\ast}\right) \\
&  +\sum_{m=M+1}^{M+N}\left(  \left\langle a_{m}a_{n}\right\rangle _{0}%
V_{km}U_{jn}^{\ast}+\text{c.c.}\right) \nonumber\\
B_{kj}^{\left(  th\right)  }  &  =\sum_{m=M+1}^{M+N}\left\langle
a_{m}^{\dagger}a_{n}\right\rangle _{0}\left(  U_{kn}V_{jm}+V_{km}U_{jn}^{\ast
}\right) \nonumber\\
&  +\sum_{m=M+1}^{M+N}\left(  \left\langle a_{m}a_{n}\right\rangle _{0}%
V_{km}V_{jn}+\text{c.c.}\right)  \;.
\end{align}
Both $A$ and $B$ form complex $M\times M$ matrices; however, $A$ must be
Hermitian, while $B$ is not. This represents an additional $3M^{2}$ real
parameters, giving a total of $7M^{2}$ that completely specifies a given
Gaussian evolution map (so called because, if the initial state of the system
is Gaussian, it will remain Gaussian).

The functions $A_{kj}^{\left(  0\right)  }$ and $B_{kj}^{\left(  0\right)  }$
represent the solution for a zero-temperature reservoir; therefore, they
represent the quantum, or zero-point fluctuations. The functions
$A_{kj}^{\left(  th\right)  }$ and $B_{kj}^{\left(  th\right)  }$ represent
the thermal fluctuations (when the reservoir is assumed to be in a thermal
state), and other effects that may arise due to, e.g., squeezing in the
reservoir modes.

\section{Single-mode Dynamics}

The above result may be written in a simpler fashion for the case of a single
oscillator taken as the system of interest:
\end{subequations}
\begin{align}
\chi\left(  \beta,t\right)  =  &  \chi\left(  U\beta-V\beta^{\ast},0\right)
\nonumber\\
&  \times\exp\left(  -A\left\vert \beta\right\vert ^{2}+\frac{1}{2}B\beta
^{2}+\frac{1}{2}B^{\ast}\beta^{\ast2}\right)  \;, \label{eq:solution}%
\end{align}
where the indices $1,1$ are dropped. The single-mode Gaussian map is
completely characterized by $7$ real parameters (since $A$ is real, and $U$,
$V$ and $B$ are complex).

\begin{figure}[ptb]
\begin{center}
\includegraphics[
height=8.0cm,
width=10.0cm
]%
{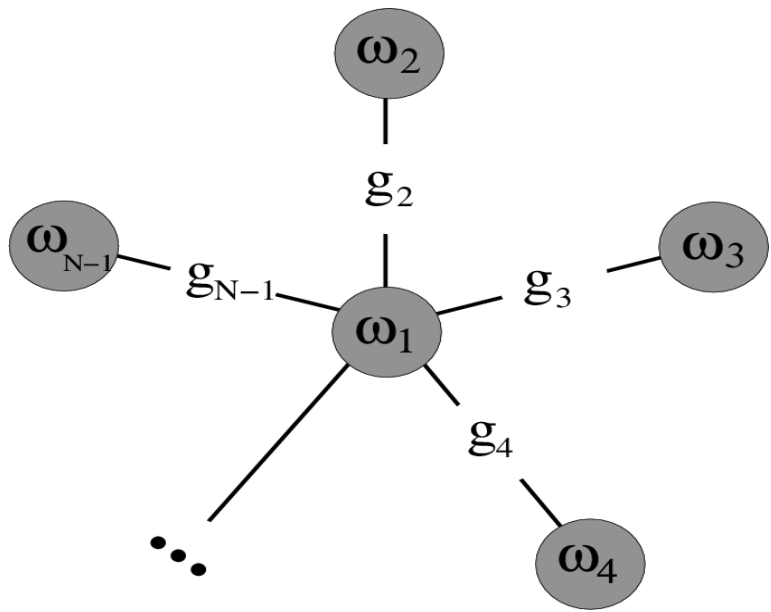}%
\caption{The system of
interest (represented by a single harmonic oscillator of the original network)
interacting with the normal modes of the diagonalized reservoir (represented
by the remaining oscillators of the network).}%
\label{fig:fig2}%
\end{center}
\end{figure}

When a single mode is considered as the system of interest, we can perform a
diagonalization of the reservoir part of the Hamiltonian, and consider the
interaction of the system with each of the reservoir normal modes, as depicted
in Fig. \ref{fig:fig2} (normal modes of the reservoir do not interact with
each other, but interact with the system).

In order to get physical results in the limit $N\rightarrow\infty$, it is
essential to keep track of the oscillator masses ($m_{k}$ in Eq.
(\ref{eq:hamiltonqp})). Essentially, the central oscillator must be much more
massive than the reservoir modes. This is the case with Brownian motion, where
the observed particle, though mesoscopic, is still much larger than the bath
of fluid molecules it interacts with. It is also the case in Quantum Optics,
where the mode inside a cavity has a much smaller mode volume (i.e., it is
concentrated in a small region) than the vacuum modes outside the cavity. We
shall consider then that the central oscillator has mass $M$ and the reservoir
modes have mass $\mu$, with $M\gg\mu$, and the renormalized frequencies and
couplings are
\begin{subequations}
\begin{align}
\omega_{1}  &  =\sqrt{\varpi_{1}^{2}+\frac{1}{M}\sum_{j=2}^{N+1}\lambda_{1j}%
}\\
\omega_{j}  &  =\sqrt{\varpi_{j}^{2}+\frac{1}{\mu}\lambda_{1j}}\quad\left(
2\leq j\leq N+1\right) \\
g_{j}  &  =\frac{1}{2\sqrt{\mu M}}\frac{\lambda_{1j}}{\sqrt{\omega_{1}%
\omega_{j}}}\quad\left(  2\leq j\leq N+1\right)
\end{align}

Dropping the first index, Eqs.(\ref{s1},\ref{s2}) become
\end{subequations}
\begin{subequations}
\begin{align}
\frac{dU_{1}}{dt}  &  =i\omega_{1}U_{1}-i\sum_{j=2}^{N}g_{j}\left(
U_{j}-V_{j}\right) \\
\frac{dV_{1}}{dt}  &  =-i\omega_{1}V_{1}-i\sum_{j=2}^{N}g_{j}\left(
U_{j}-V_{j}\right) \\
\frac{dU_{j}}{dt}  &  =i\omega_{j}U_{j}-ig_{j}\left(  U_{1}-V_{1}\right)
\quad\quad\left(  j\neq1\right) \\
\frac{dV_{j}}{dt}  &  =-i\omega_{j}V_{j}-ig_{j}\left(  U_{1}-V_{1}\right)
\quad\quad\left(  j\neq1\right)  \;.
\end{align}

The bottom two equations can be solved by considering $U_{1}$ and $V_{1}$ as
external parameters.
%(their solutions are necessary to compute $P$ and $Q$ according to \ref{eq:pqzero} and \ref{eq:pqtemp}).
Then, by substituting them into the top two equations, we get a pair of
coupled integro-differential equations:
\end{subequations}
\begin{subequations}
\begin{align}
\frac{dU_{1}}{dt}  &  =i\omega_{1}U_{1}+i\int_{0}^{t}d\tau h\left(
t-\tau\right)  \left(  U_{1}\left(  \tau\right)  -V_{1}\left(  \tau\right)
\right) \label{eq:u1integro}\\
\frac{dV_{1}}{dt}  &  =-i\omega_{1}V_{1}+i\int_{0}^{t}d\tau h\left(
t-\tau\right)  \left(  U_{1}\left(  \tau\right)  -V_{1}\left(  \tau\right)
\right)  \;, \label{eq:v1integro}%
\end{align}
which depends on the reservoir topology only through the function
\end{subequations}
\begin{equation}
h\left(  t\right)  =\sum_{j=2}^{N+1}g_{j}^{2}\sin\left(  \omega_{j}t\right)
=\frac{1}{4\mu M\omega_{1}}\sum_{j=2}^{N+1}\frac{\lambda_{j}^{2}}{\omega_{j}%
}\sin\left(  \omega_{j}t\right)  \;,
\end{equation}
which in turn is related to the Fourier transform of the reservoir spectral
density
\begin{equation}
J\left(  \omega\right)  =\sum_{j=2}^{N+1}g_{j}^{2}\delta\left(  \omega
-\omega_{j}\right)  =\frac{1}{4\mu M\omega_{1}}\sum_{j=2}^{N+1}\frac
{\lambda_{j}^{2}}{\omega_{j}}\delta\left(  \omega-\omega_{j}\right)
\end{equation}

This is the homogeneous part of the solution. To obtain the inhomogeneous one,
we need to use the solution found previously for $U_{k}$ and $V_{k}$ in terms
of the now known $U_{1}$ and $V_{1}$, and then use Eqs. (\ref{eq:pqzero}) and
(\ref{eq:pqtemp}).

\section{Master Equation}

The complete solution for single-mode dynamics is Eq. (\ref{eq:solution}),
with time-dependent functions $U$, $V$, $A$ and $B$. It was derived by
assuming an explicit microscopic model for the reservoir as a set of other
modes, which are coupled to the mode of interest, but over which the
experimenter has little control (except for macroscopic parameters such as
temperature). In this section, our goal is to find a dynamical equation (in
fact, a master equation) whose solution is precisely Eq. (\ref{eq:solution}),
but which does not need to involve any other degrees of freedom, besides those
of the system.

We start by differentiating Eq. (\ref{eq:solution}) with respect to time, and
then mapping it from phase space back to Hilbert space:%
\begin{equation}
\frac{d\rho}{dt}=-i\left[  H_{S}\left(  t\right)  ,\rho\left(  t\right)
\right]  +\mathcal{D}_{t}\left(  \rho\left(  t\right)  \right)  ,
\label{eq:master}%
\end{equation}
where we have a time-dependent effective Hamiltonian
\begin{equation}
H_{S}\left(  t\right)  =\omega\left(  t\right)  a^{\dagger}a+\xi\left(
t\right)  a^{\dagger2}+\xi^{\ast}\left(  t\right)  a^{2}\;,
\label{eq:masterham}%
\end{equation}
and a time-dependent dissipation super-operator,
\begin{align}
\mathcal{D}_{t}\left(  \rho\right)  =  &  \frac{\gamma_{1}\left(  t\right)
+\gamma_{2}\left(  t\right)  }{2}\left(  \left[  a\rho,a^{\dagger}\right]
+\left[  a,\rho a^{\dagger}\right]  \right) \nonumber\\
&  +\frac{\gamma_{2}\left(  t\right)  }{2}\left(  \left[  a^{\dagger}%
\rho,a\right]  +\left[  a^{\dagger},\rho a\right]  \right) \nonumber\\
&  -\frac{1}{2}\left(  \eta\left(  t\right)  \left(  \left[  a^{\dagger}%
\rho,a^{\dagger}\right]  +\left[  a^{\dagger},\rho a^{\dagger}\right]
\right)  +\text{H.c.}\right)  \;. \label{eq:masterdiss}%
\end{align}

This master equation depends on $7$ real time-dependent parameters, which in
turn depend on the $7$ real parameters that define solution
Eq.(\ref{eq:solution}); the three real parameters
\begin{subequations}
\begin{equation}
\omega\left(  t\right)  =\frac{1}{\left\vert U\right\vert ^{2}-\left\vert
V\right\vert ^{2}}\Im\left(  U^{\ast}\frac{dU}{dt}-V^{\ast}\frac{dV}%
{dt}\right)  \;,
\end{equation}%
\begin{align}
\gamma_{1}\left(  t\right)  =  &  \frac{-2}{\left\vert U\right\vert
^{2}-\left\vert V\right\vert ^{2}}\Re\left(  U^{\ast}\frac{dU}{dt}-V^{\ast
}\frac{dV}{dt}\right) \nonumber\\
=  &  -\frac{d}{dt}\log\left(  \left\vert U\right\vert ^{2}-\left\vert
V\right\vert ^{2}\right)  \;, \label{eq:gammafrommapa}%
\end{align}%
\begin{equation}
\gamma_{2}\left(  t\right)  =\frac{dA}{dt}+\gamma_{1}\left(  A-\frac{1}%
{2}\right)  +2\Im\left(  \xi^{\ast}B\right)  \;, \label{eq:gamma2}%
\end{equation}
and the two complex parameters
\begin{equation}
\xi\left(  t\right)  =\frac{-i}{\left\vert U\right\vert ^{2}-\left\vert
V\right\vert ^{2}}\left(  U\frac{dV}{dt}-V\frac{dU}{dt}\right)  ,
\end{equation}%
\begin{equation}
\eta\left(  t\right)  =\frac{dB}{dt}+\left(  \gamma_{1}+2i\omega\right)
B+2i\xi A. \label{eq:eta}%
\end{equation}
The time-dependent functions $\omega\left(  t\right)  $, $\gamma_{1}\left(
t\right)  $ and $\xi\left(  t\right)  $ are independent of the initial state
of the reservoir, while $\gamma_{2}\left(  t\right)  $ and $\eta\left(
t\right)  $ depend on it.

The dissipator, Eq. (\ref{eq:masterdiss}), is not explicitly in Lindblad-like
form, but can be put into it,
\end{subequations}
\begin{equation}
\mathcal{D}_{t}\left(  \rho\right)  =\sum_{n=1}^{2}\frac{\lambda_{n}\left(
t\right)  }{2}\left(  \left[  L_{n}\left(  t\right)  \rho,L_{n}^{\dagger
}\left(  t\right)  \right]  +\left[  L_{n}\left(  t\right)  ,\rho
L_{n}^{\dagger}\left(  t\right)  \right]  \right)  \label{eq:masterdisslind}%
\end{equation}
by defining the Lindblad operators
\begin{subequations}
\begin{align}
L_{1}\left(  t\right)   &  =\cos\left(  \frac{\theta}{2}\right)  a-\sin\left(
\frac{\theta}{2}\right)  \frac{\eta}{\left\vert \eta\right\vert }a^{\dagger
}\label{1}\\
L_{2}\left(  t\right)   &  =\cos\left(  \frac{\theta}{2}\right)  a^{\dagger
}+\sin\left(  \frac{\theta}{2}\right)  \frac{\eta^{\ast}}{\left\vert
\eta\right\vert }a\;, \label{2}%
\end{align}
and Lindblad rates
\end{subequations}
\begin{subequations}
\begin{align}
\lambda_{1}\left(  t\right)   &  =\frac{\gamma_{1}}{2}+\frac{\gamma_{1}%
}{\left\vert \gamma_{1}\right\vert }\sqrt{\frac{\gamma_{1}^{2}}{4}+\left\vert
\eta\right\vert ^{2}}+\gamma_{2}\\
\lambda_{2}\left(  t\right)   &  =\frac{\gamma_{1}}{2}-\frac{\gamma_{1}%
}{\left\vert \gamma_{1}\right\vert }\sqrt{\frac{\gamma_{1}^{2}}{4}+\left\vert
\eta\right\vert ^{2}}+\gamma_{2}\;,
\end{align}
with the auxiliary definition
\end{subequations}
\begin{equation}
\theta=\arctan\left(  \frac{2\left\vert \eta\right\vert }{\gamma_{1}}\right)
\quad\left(  -\frac{\pi}{2}\leq\theta\leq\frac{\pi}{2}\right)
\end{equation}

The standard master equation derived with the Born-Markov approximation has
the same form as equations Eq. (\ref{eq:master})-(\ref{eq:masterdiss}), but
with constant-in-time parameters. In it, each term has a physical meaning:

\begin{itemize}
\item The first term in Eq. (\ref{eq:masterham}), with $\omega\left(
t\right)  =\omega_{1}+\Delta\omega\left(  t\right)  $, accounts for the free
dynamics of the system, modified by a frequency shift due to its interaction
with the reservoir.

\item The second term in Eq. (\ref{eq:masterham}) is a squeezing term, arising
from an asymmetry between position and momentum variables in the coupling
Hamiltonian. However, in the weak-coupling regime, this term is small (being
exactly zero in the RWA), leading to a negligible squeezing effect.

\item $\gamma_{1}\left(  t\right)  $ is a decay rate,
%of an initial excitation present in the system.
that drives the center of the system wave-packet towards its equilibrium at
the origin of phase space.

\item $\gamma_{2}\left(  t\right)  $ is a diffusion coefficient, related to
injection of extra noise into the system due to non-zero reservoir temperature
and counter-rotating terms, which only spreads the wave-packet without
affecting the trajectory of its center.

\item $\eta\left(  t\right)  $ is a coefficient of anomalous diffusion, which
injects different levels of noise in position and momentum. From Eqs.
(\ref{1},\ref{2}), we see that, when $\eta\neq0$, the Lindblad operators are
not given by $a$ and $a^{\dagger}$, but by linear combinations of the two,
giving rise to anomalous diffusion.
\end{itemize}

\subsection{Markovian and non-Markovian behavior}

An interesting discussion in the current literature (see Ref.
\cite{NonMarkovian} and references therein) concerns non-Markovian behavior.
The Born-Markov approximation always leads to a Lindblad equation with a
dissipator written in the form of Eq.(\ref{eq:masterdisslind}), with rates
$\lambda_{n}\left(  t\right)  $, which are positive but may vary in time (in
which case it can be called a \emph{time-dependent Markovian process}). If, at
any given time, one of these rates assumes a negative value, then it is said
to be a \emph{non-Markovian process}, according to
%$Henderson-Vedral \cite{Vedral}.
the divisibility criterion of Rivas-Huelga-Plenio \cite{NonMarkovian,RHP}.

The model we have developed allows us to compute these rates exactly from the
solution, obtained through the system-reservoir interaction Hamiltonian. We
can thus describe the system as \emph{Markovian} if the following conditions
hold for all times $t$:%
\begin{subequations}
\begin{align}
\gamma_{1}\left(  t\right)  +2\gamma_{2}\left(  t\right)   &  \geq0\\
\gamma_{1}\left(  t\right)  \gamma_{2}\left(  t\right)  +\gamma_{2}^{2}\left(
t\right)  -\left\vert \eta\left(  t\right)  \right\vert ^{2}  &  \geq0\;,
\end{align}
where the functions are defined in Eq. (\ref{eq:gammafrommapa}), Eq.
(\ref{eq:gamma2}) and Eq. (\ref{eq:eta}).

\section{Rotating Wave Approximation}

In many physical systems described by the Hamiltonian of Eq.
(\ref{eq:hamiltona}), the typical coupling intensity, $\left\vert
g_{kj}\right\vert $, is many orders of magnitude smaller than the frequencies
$\omega_{k}$, characterizing the \emph{weak coupling regime}. It is then a
good approximation to drop the counter-rotating terms ($a_{k}a_{j}$ and
$a_{k}^{\dagger}a_{j}^{\dagger}$), a procedure which is known as the
\emph{rotating wave approximation} (\emph{RWA}). Eqs. (\ref{s1},\ref{s2}) are
greatly simplified,
%as they become a system of only $N+M$ coupled equations,
with $V_{kj}=0$ and $U_{kj}$ obeying:
\end{subequations}
\begin{equation}
\frac{dU_{kj}}{dt}=i\omega_{j}U_{kj}-i\sum_{n=1}^{N}U_{kn}g_{nj}\;.
\end{equation}

The condition $V_{kj}=0$ (for all $kj$) implies both $\xi\left(  t\right)  =0$
(no squeezing term in the effective system Hamiltonian) and $B^{\left(
0\right)  }=0$ and, unless the reservoir initial state has some degree of
squeezing (i.e., $\left\langle a_{m}a_{n}\right\rangle _{0}\neq0$ for some
$m,n$), then also $B^{\left(  th\right)  }=0$. Together, this implies that
$\eta\left(  t\right)  =0$. The condition $\xi\left(  t\right)  =\eta\left(
t\right)  =0$ is required to maintain the symmetry between position and
momentum variables (the exchange $\left( \hat{q},\hat{p}\right)
\leftrightarrow\left( \hat{p},-\hat{q}\right) $ leaves the RWA Hamiltonian
unchanged, while it changes the one in Eq. (\ref{eq:hamiltonqp})). Therefore,
in RWA, the squeezing term in Eq. (\ref{eq:masterham}) and the last term in
Eq. (\ref{eq:masterdiss}) both vanish at all times, leading to the usual three
terms (frequency shift, dissipation and diffusion) in the expression. The
Markovianity condition is then simplified to
\begin{subequations}
\begin{align}
\gamma_{1}\left(  t\right)  +2\gamma_{2}\left(  t\right)   &  \geq0\\
\gamma_{2}\left(  t\right)   &  \geq0
\end{align}

\section{Natural Basis For System Evolution}

It is a well known result \cite{GlauberBook} that a coherent state remains
coherent when in contact with a reservoir at absolute zero, if one assumes
RWA. This makes coherent states a natural basis to analyze system dynamics,
ultimately motivating Glauber and Sudarshan to define the normal-order
quasi-probability $P$ function:
\end{subequations}
\begin{equation}
\rho\left(  t\right)  =\int d^{2M}\left\{  \alpha\right\}  P\left(  \left\{
\alpha\right\}  ,t\right)  \left\vert \left\{  \alpha\right\}  \right\rangle
\left\langle \left\{  \alpha\right\}  \right\vert .
\end{equation}

{We have returned to the general case, where the system is composed of $M$
modes.} The coherent state follows a dynamics in phase space that can be
written $\left\vert \left\{  \alpha\right\}  \right\rangle \rightarrow
\left\vert \left\{  \alpha\left(  t\right)  \right\}  \right\rangle $, where
$\left\{  \alpha\left(  t\right)  \right\}  $ is given by (compare with Eq.
(\ref{eq:linear}))
\begin{equation}
\alpha_{k}\left(  t\right)  =\sum_{j=1}^{M}\left(  U_{kj}\alpha_{j}%
+V_{kj}\alpha_{j}^{\ast}\right)  \quad\left(  1\leq k\leq M\right)  \;.
\label{eq:lineardirect}%
\end{equation}
Combining these two equations, we have the familiar result
\begin{equation}
\rho\left(  t\right)  =\int d^{2M}\left\{  \alpha\right\}  P\left(  \left\{
\alpha\right\}  ,0\right)  \left\vert \left\{  \alpha\left(  t\right)
\right\}  \right\rangle \left\langle \left\{  \alpha\left(  t\right)
\right\}  \right\vert . \label{eq:glauberevolution}%
\end{equation}

The fact that coherent states remain coherent is intimately connected with the
fact that the vacuum is a stationary state of this non-unitary evolution.
However, for non-zero temperature, or when one includes the counter-rotating
terms, this is no longer true: coherent states do not maintain their
coherence, and we must resort to another basis, formed by Gaussian states. In
the same way that the coherent states are generated by displacing the vacuum,
the time-dependent Gaussian basis states are generated by displacing a
squeezed thermal state:
\begin{equation}
\rho_{B}\left(  \left\{  \alpha\right\}  ,t\right)  =D\left(  \left\{
\alpha\right\}  \right)  \rho_{o}\left(  t\right)  D^{\dagger}\left(  \left\{
\alpha\right\}  \right)  ,
\end{equation}
where $\rho_{o}\left(  t\right)  $ is obtained by allowing an initial vacuum
state to evolve in accordance with the solution presented in Eq.
(\ref{eq:solution}):
\begin{equation}
\left\vert 0\right\rangle \left\langle 0\right\vert \rightarrow\rho_{o}\left(
t\right)  =\int d^{2M}\left\{  \alpha\right\}  P_{o}\left(  \left\{
\alpha\right\}  ,t\right)  \left\vert \left\{  \alpha\right\}  \right\rangle
\left\langle \left\{  \alpha\right\}  \right\vert \label{eq:evolvacuum}%
\end{equation}

Adopting then this natural Gaussian basis, we can write the evolution of any
initial state as:
\begin{equation}
\rho\left(  t\right)  =\int d^{2M}\left\{  \alpha\right\}  P\left(  \left\{
\alpha\right\}  ,0\right)  \rho_{B}\left(  \left\{  \alpha\left(  t\right)
\right\}  ,t\right)  . \label{eq:evolany}%
\end{equation}

Combining Eq. (\ref{eq:evolvacuum}) and Eq. (\ref{eq:evolany}), we can rewrite
the evolution of an arbitrary initial state (albeit one with a reasonably
well-defined $P$ function) as
\begin{align}
\rho\left(  t\right)  =  &  \int d^{2M}\left\{  \alpha\right\}  \int
d^{2M}\left\{  \eta\right\}  P\left(  \left\{  \alpha\right\}  ,0\right)
P_{o}\left(  \left\{  \eta\right\}  ,t\right) \nonumber\\
&  \times\left\vert \left\{  \eta+\alpha\left(  t\right)  \right\}
\right\rangle \left\langle \left\{  \eta+\alpha\left(  t\right)  \right\}
\right\vert , \label{eq:naturalbasis}%
\end{align}
where $\left\{  \alpha\left(  t\right)  \right\}  $ describe the evolution of
the \emph{center} of the wavepacket (which obeys a classical equation of
motion, as required by the Ehrenfest theorem, and is independent of the state
of the reservoir) and $P_{o}\left(  \left\{  \eta\right\}  ,t\right)  $
describe the evolution of the \emph{shape} of the wavepacket.

When the RWA and an absolute-zero reservoir are assumed, the wavepacket is not
distorted, and $P_{o}\left(  \left\{  \eta\right\}  ,t\right)  $ reduces to a
delta function at the origin, making Eq. (\ref{eq:naturalbasis}) identical to
Eq. (\ref{eq:glauberevolution}). Therefore, Eq. (\ref{eq:naturalbasis}) is a
generalization of Eq. (\ref{eq:glauberevolution}) and we have obtained a
generalization of the dynamics described in Ref. \cite{GlauberBook}.

Another way to look at this result is that the displaced phase-space
quasi-probability function is convoluted with another function, which accounts
for the change in shape.
\begin{equation}
P\left(  \left\{  \alpha\right\}  ,t\right)  =\int d^{2M}\left\{
\gamma\right\}  P\left(  \left\{  \gamma\right\}  ,0\right)  P_{o}\left(
\left\{  \alpha-\gamma\left(  t\right)  \right\}  ,t\right)
\end{equation}
For a single mode, the center path follows $\alpha\left(  t\right)
=U_{1}\alpha+V_{1}\alpha^{\ast}$, $U_{1}$ and $V_{1}$ being given by the
solutions to Eqs. (\ref{eq:u1integro}) and (\ref{eq:v1integro}). The function
$P_{o}\left( \left\{  \alpha\right\}  ,t\right) $ is just the solution when
the initial state is the vacuum, i.e., it satisfies the initial condition
$P_{o}\left( \left\{  \alpha\right\}  ,0\right)  = \delta^{\left( 2\right)
}\left( \alpha\right) $. Under the RWA, this continues to be true at all
times, $P_{o}^{\text{RWA}}\left( \left\{  \alpha\right\}  ,t\right)  =
\delta^{\left( 2\right) }\left( \alpha\right) $.

\section{Conclusions}

We have presented a technique to derive an exact master equation for the
system-reservoir dynamics under the strong coupling regime, where neither the
rotating-wave-approximation nor the secular approximation apply. To this end,
we adopted the strategy of considering a network of bosonic systems coupled to
each other, picking out one of them as the system of interest and leaving the
rest to play the role of the reservoir. Working {with phase-space distribution
functions and Gaussian states, we generalize an earlier result by Glauber,
that a coherent state remains coherent despite dissipation when coupled to a
zero temperature reservoir. We demonstrate that t}here is a class of Gaussian
states which serves as a generalization of the coherent state basis of the
Glauber-Sudarshan $P$ representation. This class of Gaussian states follows
from the distortion of the vacuum state which, in the strong-coupling regime,
is no longer a stationary state, even for a zero temperature reservoir. We
have also presented an investigation of the conditions that lead to a
non-completely-divisible map, and thus non-Markovian dynamics. So far,
conditions for non-Markovianity have been studied for finite Hilbert spaces
under the rotating-wave and/or secular approximations. We remark that a master
equation similar to the one derived here has been obtained using the Path
Integrals approach \cite{HPZ}. The simplicity of our development, using
phase-space distribution functions, offers the significant advantage of
enabling us to cast the problem as the solution of a linear system of equations.

%\begin{thebibliography}{99}                                                                                               %
%\end{thebibliography}

%{\Large \textbf{Acknowledgements}}

\begin{acknowledgments}
%{\textbf{Acknowledgements}}

The authors acknowledge financial support from PRP/USP within the Research
Support Center Initiative (NAP Q-NANO) and FAPESP, CNPQ and CAPES, Brazilian agencies.
\end{acknowledgments}

%\textbf{Figure Captions}
%Fig. 1 Network of coupled quantum harmonic oscillators in a general topology.
%Fig. 2 The system of interest (represented by a single harmonic oscillator of
%the original network) interacting with the normal modes of the diagonalized
%reservoir (represented by the remaining oscillators of the network).

\end{document}